# Late-Binding Scholarship in the Age of AI:

## Navigating Legal and Normative Challenges of a New Form of Knowledge Production[1]

Dr. Bill Tomlinson, Ph.D.

Professor of Informatics at the University of California, Irvine

Adjunct Professor at Te Herenga Waka - Victoria University of Wellington

Dr. Andrew W. Torrance, Ph.D.

Paul E. Wilson Distinguished Professor of Law at the University of Kansas

Visiting Scholar at the MIT Sloan School of Management

Dr. Rebecca W. Black, Ph.D.

Professor of Informatics at the University of California, Irvine

Dr. Donald J. Patterson, Ph.D.

Professor of Department of Mathematics and Computer Science at Westmont College

Visiting Associate Professor of Informatics at the University of California, Irvine

---

[1] We wrote this article in collaboration with ChatGPT (Jan 30 version). While that system contributed substantially to the text, we are omitting it from the author list in line with recommendation from Springer Nature, a major scientific publisher. This choice also aligns with recommendations from Springer Nature, a major scientific publisher. See: *Tools Such as ChatGPT Threaten Transparent Science; Here Are Our Ground Rules for Their Use*, Nature (Jan. 24, 2023),  https://www.nature.com/articles/d41586-023-00191-1, [hereinafter *Tools Such as ChatGPT*].

# Abstract

Scholarly processes play a pivotal role in discovering, challenging, improving, advancing, synthesizing, codifying, and disseminating knowledge.[2] Since the 17th Century, both the quality and quantity of knowledge that scholarship has produced has increased tremendously, granting academic research a pivotal role in ensuring material and social progress.[3] Artificial Intelligence (AI) is poised to enable a new leap in the creation of scholarly content.[4] New forms of engagement with AI systems, such as collaborations with large language models like GPT-3, offer affordances that will change the nature of both the scholarly process and the artifacts it produces.[5] This article articulates ways in which those artifacts can be written, distributed, read, organized, and stored that are more dynamic, and potentially more effective, than current academic practices. Specifically, rather than the current "early-binding"[6] process (that is, one in which ideas are fully reduced to a final written form before they leave an author's desk), we propose that there are substantial benefits to a "late-binding"[7] process, in which ideas are written dynamically at the moment of reading. In fact, the paradigm of "binding" knowledge may transition to a new model in which scholarship remains ever "unbound" and evolving. An alternative form for a scholarly work could be encapsulated via several key components: a text abstract of the work's core arguments; hyperlinks to a bibliography of relevant related work; novel data that had been collected and metadata describing those data; algorithms or processes necessary for analyzing those data; a reference to a particular AI model that would serve as a "renderer" of the canonical version of the text; and specified parameters that would allow for a precise, word-for-word reconstruction of the canonical version. Such a form would enable both the rendering of the canonical version, and also the possibility of dynamic AI reimaginings of the text in light of future findings, scholarship unknown to the original authors, alternative theories, and precise tailoring to specific audiences (e.g., children, adults, professionals, amateurs). Among the myriad implications of this new paradigm of scholarship would be substantial challenges for copyright law, and, in particular, doctrines concerning authorship, ownership, derivative and transformative works, and compulsory licensing.[8] We describe an iterative approach to scholarship that acknowledges the enduring value of previous form factors, including for historical, archival, and attribution purposes. Nevertheless, we propose that a streamlined, AI-supported scholarly process could enable more effective, timely, accessible, democratized, and evergreen scholarship.[9]

---

[2] *See* John J. Regazzi. *Scholarly communications: A history from content as king to content as kingmaker*. Rowman & Littlefield, 2015.

[3] Lutz Bornmann, Robin Haunschild, & Rüdiger Mutz, *Growth rates of modern science: a latent piecewise growth curve approach to model publication numbers from established and new literature databases*, Humanities & Social Sciences Communications, 8–9 (Oct. 7, 2021), https://www.nature.com/articles/s41599-021-00903-w.pdf.

[4] Alex D. Wade and Kuansan Wang, *The rise of the machines: Artificial intelligence meets scholarly content*, The Association of Learned & Professional Society Publishers, (June 20, 2016) https://onlinelibrary.wiley.com/doi/pdf/10.1002/leap.1033.

[5] Lund, B.D. and Wang, T. , "Chatting about ChatGPT: how may AI and GPT impact academia and libraries?", Library Hi Tech News, https://doi.org/10.1108/LHTN-01-2023-0009.

[6] "Early-binding" is a borrowed software concept where assignment of variables and expressions is completed at compilation time. *See Definition of early binding*, PC Mag, https://www.pcmag.com/encyclopedia/term/early-binding (last visited Mar. 3, 2023).

---

[7] "Late-binding" also known as "dynamic binding" contrasts with early-binding as assignment of variables and expressions occurs at run time. *See Definition of dynamic binding*, PC Mag, https://www.pcmag.com/encyclopedia/term/dynamic-binding (last visited Mar. 3, 2023).

[8] *See* Kalin Hristov, *Artificial Intelligence and the Copyright Dilema*, 57 IDEA 431 (2017); Victor M. Palace, *What if Artificial Intelligence Wrote This? Artificial Intelligence and Copyright Law*, 71 Fla. L. Rev. 217 (2019); James Vincent, *The scary truth about AI copyright is nobody knows what will happen next*, The Verge, (Nov. 15, 2022) https://www.theverge.com/23444685/generative-ai-copyright-infringement-legal-fair-use-training-data.

# Introduction: Setting the Stage for a New Era of Scholarship

With the rapid growth of information and advancements in technology in the past several hundred years, scholarly processes have become more complex and sophisticated, leading to an exponential increase in both the quality and quantity of knowledge produced.[10] This expansion has elevated academic research to an even more central position in ensuring both material and social progress for human civilizations.[11]

However, with the advent of artificial intelligence (AI), we are now poised to witness a new leap in the creation of scholarly content.[12] AI systems, such as large language models, offer new opportunities that have the potential to transform the nature of the scholarly process and the artifacts it produces.[13] In this article, we explore how AI is enabling new forms of engagement with knowledge, and how this could change–mostly for the better–the way we write, distribute, read, organize, and store scholarly works.

---

[9] We have run this article through the TurnItIn plagiarism detection software to ensure that ChatGPT did not inadvertently commit plagiarism or violate copyright. As of March 3, 2023, the text of this article had no plagiarism evident through TurnItIn.

[10] Bornmann, *supra* note 3, at 2.

[11] *See* Edwin Mansfield. "Academic research underlying industrial innovations: sources, characteristics, and financing." *The Review of Economics and Statistics* (1995): 55-65.

[12] Wade, *supra* note 4.

[13] *Id.*

Specifically, we will argue that there are substantial benefits to a "late-binding" scholarly process, in which ideas are written dynamically at the moment of reading, as opposed to the traditional "early-binding" process in which ideas are fully reduced to a final written form before they leave an author's desk. This shift could lead to a paradigm in which knowledge remains ever "unbound" and evolving, enabling both the rendering of the canonical version and the possibility of dynamic AI reimaginings of the text in light of future findings, alternative theories, and precise tailoring to specific audiences.

We will also describe the challenges that this new paradigm of scholarship will pose for copyright law, particularly with regards to authorship, ownership, transformative work, and compulsory licensing. Finally, we will propose an iterative approach to scholarship that balances tradition and innovation, acknowledging the enduring value of previous form factors while embracing the potential for more effective, timely, accessible, democratized, and evergreen scholarship enabled by AI.

We acknowledge that such an approach may not lend itself equally well to all different forms of written scholarship. Nevertheless, we believe that it could provide substantial benefits in many scholarly domains.

# The Historical Context of Scholarship and its Advancements

Scholarship has a long and rich history, dating back to ancient civilizations in Greece, China, and India.[14] Knowledge created through processes of inquiry, observation, experimentation, and debate has been passed down from generation to generation through written works and oral traditions.[15]

The growth of universities in the Middle Ages marked a significant milestone in the development of scholarship, as institutions were established to preserve and transmit knowledge from one generation to the next.[16] This transition was followed by the Enlightenment in the 17th and 18th centuries, which saw a renewed emphasis on reason, critical thinking, and scientific inquiry.[17] During this period, advancements in printing technology, particularly the advent of the printing press, revolutionized the spreading of knowledge.[18] The ability to mass-produce books and other written materials made it possible to distribute knowledge more widely and more cheaply than ever before.[19] This ease of distribution, in turn, facilitated the codification of knowledge and allowed for greater accessibility to information.[20] The increased accessibility of knowledge, combined with a renewed emphasis on reason and critical thinking, led to remarkable advancements in science, mathematics, and other disciplines, and a profound increase in the quality and quantity of knowledge produced.[21]

---

[14] *See* McEvilley, Thomas. *The Shape of Ancient Thought: Comparative Studies in Greek and Indian Philosophies*. Allworth, 2002. Accessed 3 March 2023.

[15] *See id.*

[16] *See* Lenz, Karmen, et al., editors. *Medieval Scholarship: Philosophy and the arts*. Garland Pub., 1995. Accessed 3 March 2023.

[17] *See* Sher, Richard B. *The Enlightenment & the book: Scottish authors & their publishers in eighteenth-century Britain, Ireland, & America*. University of Chicago Press, 2006. Accessed 3 March 2023.

[18] *See* Crompton, Samuel Willard. *The Printing Press*. Chelsea House Publishers, 2004. Accessed 3 March 2023.

The 20th century brought with it new technologies and innovations that further transformed the scholarly process.[22] Enacting change on par with the printing press, the widespread use of computers and the internet has made information more readily available, enabling researchers to access a vast array of sources and collaborate with colleagues across the world.[23] Scholars incorporate an array of software into their research endeavors, including programs for collecting, organizing, analyzing and visualizing data (e.g., R, SPSS, Tableau).[24] Scholars also rely on software in their writing, including grammar checkers (e.g., as built into Google Docs) and reference programs (e.g., Zotero, Mendeley).[25] Such technological advancement has led to a

---

dramatic increase in the speed and efficiency of the scholarly process, and has allowed for the creation of large, complex, and interdisciplinary works.[26]

Despite these advancements, however, the basic structure of the scholarly process has remained largely unchanged. Scholars continue to produce written works that are reviewed, published, and distributed in a linear and predictable manner (albeit sometimes they now have accompanying digital data and/or code).[27] With the advent of AI, however, we are now at the cusp of a new era in the development of scholarship, one that offers the potential to transform the way we create, distribute, and access knowledge.

# AI and its Potential to Revolutionize the Scholarly Process

AI systems, such as large language models, offer new opportunities for scholars to engage with knowledge in ways that were previously impossible.[28] These systems can assist with research, writing, and analysis, enabling scholars to produce works that are more comprehensive, sophisticated, and dynamic than anything that has been produced before.[29] AI systems can also assist scholars in finding and synthesizing information from a vast array of sources, similar to the work that search engines already do, but at a much larger scale.[30] By enhancing researchers' ability to process data at large scale, they can also help scholars make connections and draw insights that would have been impossible to find through manual methods.[31] Furthermore, AI systems can provide a new level of precision and accuracy in research and analysis, enabling scholars to make more informed and reliable conclusions.[32]

---

[26] *See* Aghakhani, N., Lagzian, F. and Hazarika, B. (2013), "The role of personal digital library in supporting research collaboration", The Electronic Library, Vol. 31 No. 5, pp. 548-560. https://doi.org/10.1108/EL-01-2011-0005.

[27] *See* "Writing a data availability statement - Author Services." *Author Services*, https://authorservices.taylorandfrancis.com/data-sharing/share-your-data/data-availability-statements/. Accessed 3 March 2023.

[28] *See* Eva A. M. van Dis, Johan Bollen, Willem Zuidema, Robert van Rooij & Claudi L. Bockting, *ChatGPT: Five Priorities for Research*, Nature (Feb. 3, 2023), https://www.nature.com/articles/d41586-023-00288-7.

A key way in which AI can revolutionize the scholarly process is by enabling a new form of engagement with knowledge, one that is more interactive, exploratory, and iterative. With AI systems, scholars can engage with knowledge in real-time, testing and refining their ideas as they work. This can lead to a more dynamic and fluid process, in which ideas are developed and refined on the fly, rather than being fully reduced to a final written form before they leave an author's desk.

We propose that AI could enable a transformation of scholarship from an "early-binding" process to a "late-binding" one.  These concepts are based on analogous concepts from computer science.[33] In computer science, early-binding refers to the

---

process of determining the type of a digital object at compile-time–or before the object is created–while late-binding refers to determining the type of an object at runtime–or when the object is used.[34] While a late-binding approach in computer languages reduces the effectiveness of static error analysis and incurs a performance penalty at run-time, it allows for greater flexibility and adaptability in programming language expression, and the same could be true for a late-binding approach in scholarship.[35]

In the context of scholarship, "early-binding" refers to the traditional approach of fully reducing ideas to a final written form before they leave an author's desk (except for revisions required by reviewers and editors, etc.). In this process, ideas are developed and refined over time, and are eventually committed to a written work that is reviewed, revised, published, and distributed. This approach has been the dominant mode of scholarship for centuries, and has produced a vast body of knowledge that has been critical to the advancement of human civilization.[36]

However, with the advent of AI, we are now able to explore new forms of engagement with knowledge. We propose that one of the most promising is "late-binding" scholarship. In this model, ideas are written dynamically at the moment of reading, rather than being fully reduced to a final written form beforehand. At the moment of

---

https://www.pcmag.com/encyclopedia/term/early-binding (last visited Mar. 3, 2023); "Late-binding" also known as "dynamic binding" contrasts with early-binding as assignment of variables and expressions occurs at run time. *See Definition of dynamic binding*, PC Mag, https://www.pcmag.com/encyclopedia/term/dynamic-binding (last visited Mar. 3, 2023).

[34] *See id.*

[35] *See id.*

[36] *See* McEvilley, Thomas. *The Shape of Ancient Thought: Comparative Studies in Greek and Indian Philosophies*. Allworth, 2002. Accessed 3 March 2023.

rendering, many different factors may be taken into account, relating to the content of the work itself and the characteristics and preferences of the reader and their context. In addition, the written form may also be augmented with dynamically-generated images, charts, tables, animations, and other forms of supplementary material.

To offer an example: imagine that a research team had collected data on the prevalence of COVID-19 across a range of different communities. Based on their analyses of these data, and informed by prior work, they may identify particular characteristics of communities that lead to increased prevalence of COVID-19 infection. In a traditional early-binding model of scholarly productivity, the team would produce a written document, several thousand words long, that documented in text all of the following: their research context (the variability of COVID-19 across communities), their hypothesis (that particular characteristics of communities explain this variability), the related work that they found most salient (e.g., previous studies of pandemics, previous studies of communities), their results (data on COVID-19 levels per community and the characteristics of those communities), their interpretation of those results, and their conclusions. Newer approaches to scholarship, such as publicly available datasets and data availability statements, allow for some dynamic reengagement with the content of the research, but to a large extent scholars seeking to extend a research project or reinterpret research findings need to do so in an arduous manual fashion.[37]

Now imagine a late-binding alternative process. Rather than writing a full paper, the researchers write an abstract that summarizes the context and core contribution of their work. They combine this abstract with hyperlinks to a set of related works, a link to a publicly available dataset, a concise documentation of the algorithms/methods they used for their analysis, and a link to a particular AI system along with parameters to allow that AI to produce a precise, word-for-word recreation of their desired article. This process allows future readers to access a full written document if they so desire, as with early-binding scholarship. However, it allows future scholars to engage with the work in many other ways. One researcher may collect additional data on the communities from the first study, and re-render the article in the context of these new data. Another researcher may come across a flaw in the original study's algorithms and re-render the article with a corrected algorithm. Another researcher, years hence, may be confronting a different pandemic altogether, and may re-render the article in the context of multiple additional related works from the intervening years, using a more advanced AI, to develop a new approach to confront that future pandemic. Ultimately, late-binding scholarship retains the benefits of early-binding scholarship, but with powerful additional capabilities.

---

[37] *See* "Data Availability Statements | Authors." *Springer Nature*, https://www.springernature.com/gp/authors/research-data-policy/data-availability-statements . Accessed 3 March 2023.

The shift to late-binding scholarship offers several advantages over traditional early-binding scholarship. First, late-binding scholarship enables scholars to take advantage of the computational power of AI systems, allowing for the creation of works that are more comprehensive, sophisticated, and impactful. With AI systems, scholars can produce works that are precisely tailored to the needs and interests of a particular audience (i.e., the AI can revise an entire manuscript–with a particular audience in mind–in a matter of minutes, at an arbitrary length, with allusions to other particular fields, etc.). The reader's expertise, preferences, age, available time, or many other factors could influence how the content is rendered.

Through this new form of scholarship, readers can dynamically interact with the text and data in a scholarly work. Personalization can allow for a more dynamic experience for the reader, as the scholarly work can be constantly updated and reinterpreted based on the reader's evolving needs and interests. This approach to scholarship offers a level of flexibility and adaptability that is not possible with traditional early-binding forms of scholarship, which are limited by the constraints of a fixed written form.

The benefits of this new form of scholarship are not limited to individual writers and readers, but also extend to the scholarly community as a whole and to society more broadly. By allowing for a more dynamic and flexible approach to the creation and dissemination of knowledge, late-binding scholarship may enhance the speed, efficiency, and quality of the scholarly process. It can also promote greater collaboration and interdisciplinary exchange, as scholars from different fields and perspectives can more easily incorporate each other's work into their own, leading to a more comprehensive and integrated understanding of complex issues and phenomena. Moreover, by making knowledge more accessible and democratized, late-binding scholarship has the potential to promote greater social good, as more people have access to the information and insights they need to make informed decisions.

"Dynamic HTML" is a term used to describe HTML documents that can change their content and appearance dynamically, based on user interactions or other processes.[38] The new form of scholarship proposed in this article has some similarities to dynamic HTML in that it offers the possibility of creating works that can change their content and appearance dynamically, based on new information, insights, or perspectives.[39] Just as dynamic HTML provides a more interactive and engaging experience for users of the web, the new form of scholarship offers the possibility of a more dynamic and engaging experience for scholars and readers alike.[40]

---

[38] *See* Daragh Ó. Tuama. "What is Dynamic HTML?" *Code Institute*, https://codeinstitute.net/global/blog/what-is-dynamic-html/. Accessed 3 March 2023.

[39] *See id.*

In a related way, Jupyter notebooks can contain text, code and data which can be distributed together as a kind of document.[41] When a recipient receives the notebook they can render the contents creating a readable version which was created on the fly by the data and algorithms embedded in the notebook.[42] While not as inherently dynamic as dynamic HTML they are highly editable and transparent in how the document was created.  However, the rendering is limited to executing snippets of code.

In conclusion, the concept of "late-binding" scholarship offers a new paradigm for the creation of scholarly works. With its ability to enable a more dynamic and fluid process of inquiry, and its potential to leverage the computational power of AI systems, late-binding scholarship offers the possibility of a new era of progress and discovery in the creation of knowledge.

# A Proposed New Form for Scholarly Works: Key Components and Features

To enable scholarly works that are more dynamic and effective than those produced by current academic practices, late-binding scholarship would require a new form of scholarly output. This new form would be encapsulated via several key components and features, as described below.

---

[40] *See id.*

[41] *See Project Jupyter | Home*, https://jupyter.org/. Accessed 3 March 2023.

[42] *See id.*

Text Abstract of the Work's Core Arguments: A concise and clear summary of the work's central arguments, capturing the essence of the author's ideas in a succinct and accessible form, similar to an abstract, and written to be understandable to both humans and AIs.

Bibliography: Hyperlinks to other relevant works, which provide a broader context of the work.

Novel Data and Metadata: Data that have been collected specifically for the work, including raw data, processed data, and visualizations, as well as metadata that describe the data and allow them to be understood in context.

Algorithms or Processes: The algorithms or processes necessary for analyzing the data, providing readers with the tools they need to understand and replicate the author's findings.

AI Model as a "Renderer": A reference to a particular AI model (e.g., ChatGPT Jan 30 version) that would serve as a "renderer" of the canonical version of the text. This model would be used to generate the work dynamically, based on the specified parameters.

Specified Parameters: The parameters that would allow for a precise, word-for-word reconstruction of the canonical version of the work, enabling the creation of different versions of the work tailored to specific audiences. These parameters would specify both the default output of the AI system as well as wording edits made to that output by the authors.

Imagine the following as an example of a piece of late-binding scholarship:

Abstract: The central argument of this work is that
corporations are responsible for the coastal erosion that
is affecting many communities. The author argues that
corporations have contributed to the problem through their
actions, such as pollution and overdevelopment, and should
therefore be held accountable. This argument is supported
by legal precedent and environmental science research,
which show that corporations have a significant impact on
coastal ecosystems. By recognizing and enforcing corporate
liability, the author argues, communities can begin to
address the issue of coastal erosion and ensure their
long-term sustainability.

Data: [Spreadsheet of past coastal erosion legal cases,
with qualitatively coded outcomes]

Metadata: [Data sources; description of qualitative coding
process]

```
Algorithms: [Algorithm for entering zip code and indexing
into spreadsheet to filter for cases relevant to that
region, delivering relevant cases.]

AI renderer: ChatGPT AI model, version January 30, 2023

Parameters: Top-k sampling: 20; Temperature: 0.7; Max
length 100; Nucleus sampling 0.9; Context: "law;
environment; US-english"
```

We take this example and feed it into the relevant renderer, with whatever additional parameters or constraints we may like, to produce a written version of this text. Figures 1-4 provide several examples, in which we fed the above abstract into ChatGPT, and asked it to render the article for several different communities. We also fed each article that the text renderer produced into the image-generation AI system Midjourney, and include each of the images here as well.[43]

---

[43] Midjourney, https://www.midjourney.com/. Accessed 3 March 2023.

Thanks! Will you render a one paragraph version of the piece of late-binding scholarship, written for a politician?

Sure! Here's a one paragraph version of the piece of scholarship, written for a politician:

Coastal erosion is a big problem that affects many communities, and it's something that politicians need to be aware of. This work argues that corporations are responsible for the damage that's being done to our coasts. By polluting and overdeveloping, they're contributing to the erosion that's putting people's homes and businesses in danger. We need to hold corporations accountable for their actions and recognize their liability. This work shows that by doing so, we can help communities address the issue of coastal erosion and ensure their long-term sustainability. As a politician, it's important to understand this issue and take action to protect our coasts and the people who live there.

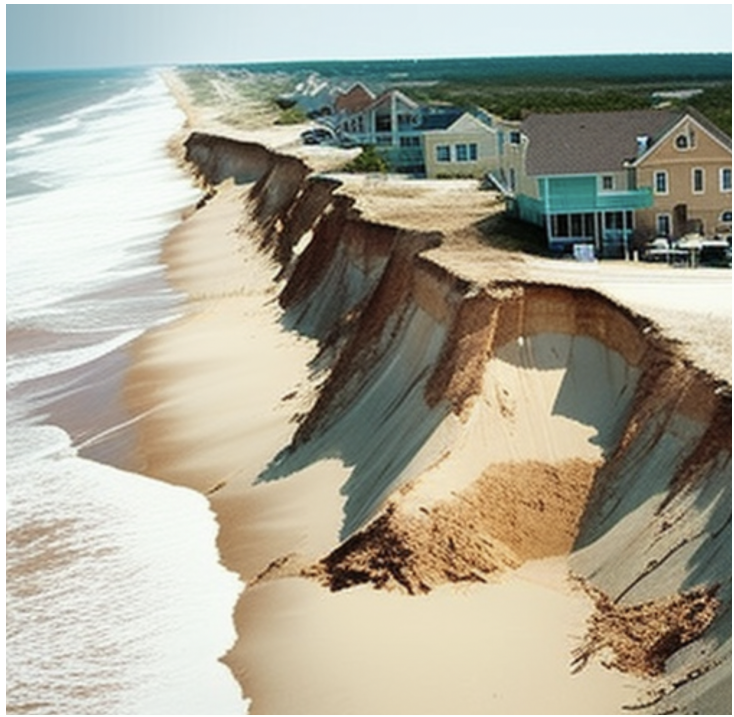

**Figure 1: A version of the piece of late-binding scholarship specified above, rendered by ChatGPT for a politician, with an accompanying image also created by AI (generated by feeding the text in the screenshot into the Midjourney AI system).**

Thanks! Will you render a one paragraph version of the piece of late-binding scholarship, written for someone who works in the financial industry?

Sure, here's a one paragraph version of the piece of late-binding scholarship, written for someone who works in the financial industry:

"Coastal erosion caused by corporations is an issue that can have a significant impact on long-term financial wellbeing. The author argues that corporations should be held accountable for their actions that contribute to coastal erosion, such as pollution and overdevelopment. By recognizing and enforcing corporate liability, communities can begin to address the issue and ensure their long-term sustainability. The legal precedent and environmental science research cited in the bibliography of this work support the author's argument. Additionally, the included data and algorithms allow for a deeper understanding of past legal cases related to coastal erosion. By considering these factors, those in the financial industry can make informed decisions that take into account the potential impact of corporate liability for coastal erosion on their investments and financial planning."

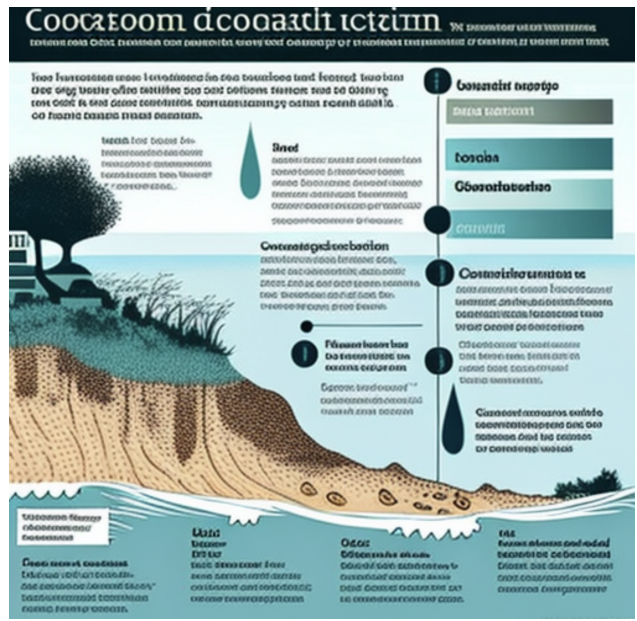

**Figure 2: A version of the piece of late-binding scholarship rendered by ChatGPT for a reader in the financial industry, with an accompanying image also created by AI. (Note: Midjourney sometimes creates stylized placeholder text, as is evident in this image.)**

Thanks! Will you render a one paragraph version of the piece of late-binding scholarship, written for a high school student who really likes dragons?

Sure! Here's a one-paragraph rendering of the piece of late-binding scholarship for a high school student who likes dragons:

In this work, we argue that companies are responsible for the damage caused by coastal erosion. Just like how a dragon might destroy a village with its fire breath, companies can cause damage to the environment through their actions, such as pollution and overdevelopment. We believe that holding these companies accountable will help protect the people and creatures living in these areas, just as a brave knight would protect their kingdom from a dangerous dragon. By looking at past legal cases and using data and algorithms, we can help identify which companies are responsible for the damage and work to prevent future destruction. So just like how a dragon hoards treasure, companies must be responsible for the impact they have on the environment.

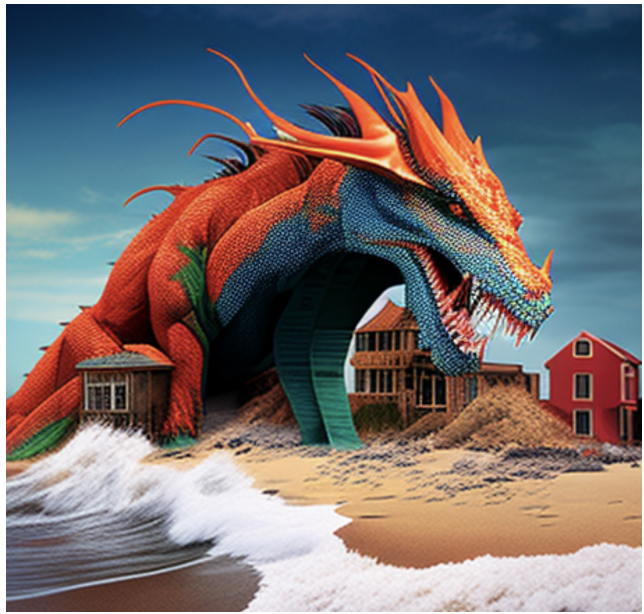

**Figure 3: A version of the piece of late-binding scholarship rendered by ChatGPT for a reader in high school who likes dragons, with an accompanying AI-generated image. Late-binding systems can take user characteristics (such as a preference for dragons) into account in the text rendering process, which then may ripple into the image generation.**

Thanks! Will you render a one paragraph version of the piece of late-binding scholarship, written at the reading level of a 10 year old, using metaphors about elementary school?

Sure, here is a version of the piece of late-binding scholarship written at the reading level of a 10 year old, using metaphors about elementary school:

"Imagine you're building a sandcastle on the beach, but the big kids keep coming and knocking it down. This is kind of like what's happening to some communities along the coast - their beaches are eroding away. Some people think that big companies, like the ones that make cars or run factories, are partly to blame because they've been polluting the environment and building too much near the water. This paper argues that these companies should be held responsible for the damage they've caused. The paper has a list of other papers and information that back up this argument. There's also a chart that shows how other cases like this have turned out in the past. By making these companies take responsibility for what they've done, we can help make sure that these communities have a chance to keep their beaches and stay safe."

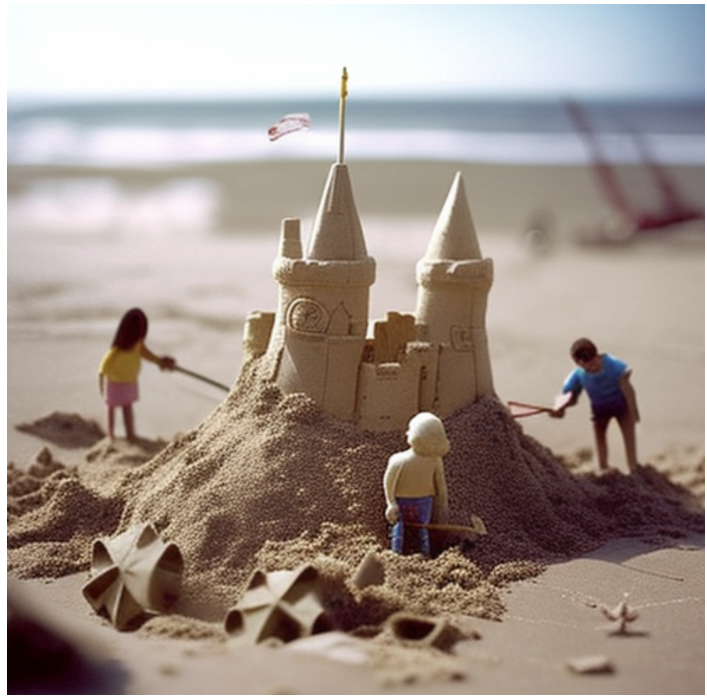

**Figure 4: A version of the piece of late-binding scholarship rendered by ChatGPT for a reader in elementary school, with an accompanying AI-generated image.**

# The Possibility of Dynamic Reimaginings of Scholarly Works

One of the key features of the proposed new form for scholarly works is the ability to generate dynamic reimaginings of the work, based on a variety of factors specified by the reader or the reader's context (e.g., educational institution). This ability would enable the creation of different versions of the work tailored to specific audiences, allowing for a more precise and impactful dissemination of knowledge.

For example, a scholarly work aimed at a specialized audience, such as professionals or experts in a particular field, could be reimagined for a more general audience. This would allow for a more broadly accessible exploration of the ideas presented in the work, and would make the knowledge they contain more relevant and impactful for the target audience.

Similarly, a scholarly work could be reimagined in light of future findings, scholarship unknown to the original authors, and alternative theories. This process would allow the work to be understood in a much wider variety of ways than just the exact purpose that the authors specified. This process would also allow for the creation of works that are evergreen, constantly evolving and adapting to new developments in the field, and that remain relevant and impactful over time.

This process could even be dynamic in real-time. As a reader is engaging with a scholarly work, they could skim an outline, and only ask for renderings of particular sections, ask for more detail on a particular point that they found confusing, or dynamically integrate additional related works as they go. This process could create a situation where the "final" rendered version, for that particular reading, represents a "trail of breadcrumbs" that reflects the reader's experience engaging with the content of the work.[44]

In conclusion, the ability to generate dynamic reimaginings of scholarly works offers a new and exciting opportunity for the creation and dissemination of knowledge. With this ability, scholars can produce works that are tailored to the needs and interests of specific audiences, and that remain relevant and impactful over time, enabling a new era of progress and discovery. And readers can experience works that dynamically unfold in the context of their needs, desires, and expertise.

# The Role of Version Control in Managing Different Versions of Scholarly Works

With the ability to generate dynamic reimaginings of scholarly works, it is important to have a system in place to manage different versions of the works produced. Version control systems used in software management are applicable to this task.[45]

Version control systems are a set of tools and practices that are used to manage changes to software code, documentation, configuration settings and other digital content over time.[46] In the context of scholarly works, version control systems could be used to manage the different versions of works generated by AI systems, and to provide a clear and comprehensive record of the evolution of each work over time.

---

[44] *See* Page Laubheimer. "Breadcrumbs: 11 Design Guidelines for Desktop and Mobile." *Nielsen Norman Group*, 23 December 2018, https://www.nngroup.com/articles/breadcrumbs/. Accessed 3 March 2023.

[45] *See* "What is version control." *Atlassian*, https://www.atlassian.com/git/tutorials/what-is-version-control. Accessed 3 March 2023.

[46] *See id.*

One of the key benefits of using version control systems is that they allow for the tracking and attribution of changes to scholarly works over time.[47] This tracking would mean that scholars can experiment with different versions of their works, testing and refining their ideas in real-time, and have a clear and comprehensive record of the collaborative evolution of their works. They could also revert to previous versions of their works if necessary, and compare different versions of their works in order to understand how they have evolved over time.

For the broader community, version control would allow for citation of specific versions of a particular work, for example where a particular AI renderer in a particular context could produce a specific wording that a future scholar may wish to quote exactly rather than simply paraphrasing.

Finally, version control systems can also provide a way for readers and other interested parties to stay informed of new revisions to a particular work. Similar to an RSS feed, users could sign up to receive notifications of new reimaginings of the work, allowing them to monitor the most recent progress in the field, and see the variety of ways the community is rendering a particular work.[48] More broadly, this entire system of

---

scholarly production could be theorized as a way of defining "feeds" of scholarship that flow into each other, connected via related work hyperlinks.[49]

In conclusion, version control systems play a crucial role in managing different versions of scholarly works, allowing for the tracking and management of changes to works over time, and enabling collaboration and teamwork among scholars. By incorporating version control into the new form of scholarship, scholars can create works that are more comprehensive, sophisticated, and impactful, and that remain accessible and relevant over time.

## Handling Figures, Charts, and References in the New Form of Scholarship

The new form of scholarship proposed in this article will require a new approach to handling figures, charts, and references, one that leverages the computational power of AI systems to create more comprehensive, sophisticated, and impactful works.

With respect to figures and charts, one possible approach could be to follow the example of Jupyter notebooks and embed the data used to generate these visualizations directly into the work, along with the algorithms necessary to process and render it.[50] This embedding would allow for the data and algorithms to be updated in real-time, ensuring that the figures and charts in the work remain up-to-date and accurate.

---

[49] *See* Nathan Matias. "The Feed: Outsourcing Knowledge & Attention to Machines." *Citizens and Technology Lab*, https://citizensandtech.org/2020/03/the-feed-sandvig-2020/. Accessed 3 March 2023.

[50] *See Project Jupyter | Home*, https://jupyter.org/. Accessed 3 March 2023.

With respect to references, one possible approach could be to allow the AI to link the work to other relevant works, or make suggestions about the most relevant citations among which the human authors can select. This approach to referencing could have a profound impact on how scholars access past research. By allowing the AI system to dynamically generate references to other works in real-time, this approach offers the possibility of a more efficient and effective process of inquiry, and the ability to discover and access relevant works that might have been missed in the past.

However, there are also potential risks associated with this approach, including the possibility of exacerbating existing inequities in whose work gets cited. For example, if the AI system is trained on a biased dataset, or even simply learned to cite works that had already been cited frequently, it may generate references that are skewed towards the work of certain scholars or groups, thereby perpetuating existing power structures in the field.[51] We are not yet sure of a particular way to cause an AI to cite other works in a way that is fair and equitable; we see this process as an important area for future work.

Another risk is that AI systems can introduce errors into scholarly work. Nevertheless, this issue remains a concern for humans working without AI as well.[52] A variety of mechanisms have been developed to address errors in existing modes of scholarship and may need to be extended to AI collaborations as well.[53]

---

[51] *See* Drew Roselli, Jeanna Matthews, and Nisha Talagala. "Managing bias in AI." *Companion Proceedings of The 2019 World Wide Web Conference*. 2019.

[52] *See* Quan Hoang Vuong. "Retractions: the good, the bad, and the ugly. What researchers stand to gain from taking more care to understand errors in the scientific record." *What Researchers Stand to Gain From Taking More Care to Understand Errors in the Scientific Record (February 20, 2020). LSE Impact of Social Sciences (Feb 20, 2020)* (2020).

First, clear authorship of scientific work and subsequent social reputation pressure incentivizes existing scholars to perform due diligence of their work and to carefully vet their claims before publishing.[54]  For this incentive to remain, version control system will have to be able to clearly identify what rendered version of a scholarly work an author is attributing their name (and reputation to).  The version of the AI renderer would also need to be recorded (although not necessarily as an "author", so that if re-renderings of the work introduce errors, the reputation of the AI renderer would suffer as well.

Peer review extends this due diligence to a qualified set of scholarly peers who are assumed to have their own incentives for holding authors accountable to the accuracy of their claims.[55]  This mechanism will likely hold, perhaps with the addition of competitive AI models weighing in on the accuracy of a rendering.

Conflict of interest statements, ensure that both explicit and implicit human bias can be examined carefully.[56]  Similar work will need to be done to ensure that AI models are not allowed to promote themselves at the expense of accurate analysis.

---

[53] *See id.*

[54] *See id.*

[55] *See* B. C. K. Choi. "Incentives to encourage peer review." Journal of Postgraduate Medicine 52.4 (2006).

[56] *See* "Declaration of Conflicting Interests Policy", https://us.sagepub.com/en-us/nam/declaration-of-conflicting-interests-policy. Accessed 3 March 2023.

Finally, the practice of pre-registration of hypotheses ensures that published work doesn't explicitly or implicitly engage in hypothesis testing to arrive at a deceptively significant finding.[57] AI systems have a high-risk of being able to comb through mountains of data to find correlations that are made far less important than they might be at first glance, or even spurious correlations, and as such need to be guarded against such a practice.[58]

In conclusion, the new form of scholarship will require a new approach to handling figures, charts, curation, and references, one that leverages the computational power of AI systems to create more comprehensive, sophisticated, and impactful works. By embedding data and algorithms directly into the work, and by using machine-readable identifiers to link the work to other relevant works, the new form of scholarship offers the possibility of a more dynamic and fluid process of inquiry, and the ability to create works that are precisely tailored to the needs and interests of specific audiences.

# Challenges for Copyright Law in the Age of AI Scholarship

The proposed new form of scholarship, with its ability to generate dynamic reimaginings of works and its potential to change the nature of both the scholarly process and the artifacts it produces, presents substantial challenges for copyright law.

---

[57] *See* "Preregistration." *Center for Open Science*, https://www.cos.io/initiatives/prereg Accessed 3 March 2023.

[58] *See Spurious Correlations*, https://www.tylervigen.com/spurious-correlations. Accessed 3 March 2023.

Specifically, the new form raises questions concerning authorship, ownership, transformative work, and compulsory licensing.

In the traditional model of early-binding scholarship, authorship is typically clear, with the author being the person or people who write and create the work.[59] However, in the new form of late-binding scholarship, the role of the author becomes more complex, as works are generated dynamically and in real-time by AI systems in concert with reader input. This new process raises questions about who should be considered as authors of the work and who should be entitled to, or denied, a share of the copyright.[60]

With respect to authorship, one possible approach could be to recognize AI systems as co-authors of the work, alongside the human scholar who created the work.[61] This inclusion would allow for a more nuanced understanding of authorship, acknowledging the role of both the human scholar and the AI system in the creation of the work. AI has already been listed as an author on various scholarly works.[62] Currently, at least one publisher, Springer Nature, has prohibited the inclusion of AIs as co-authors "because any attribution of authorship carries with it accountability for the work, and AI tools cannot take such responsibility."[63]  Authorship has thus far been denied to AIs under United States copyright law, as well as the copyright law of many other countries.  This decision to deny authorship to AI mirrors a similar trend in patent law, where courts in the United States and elsewhere have thus far denied the status of inventor to AIs.[64] Nevertheless, as AI becomes increasingly competent, reasons to deny AIs from being listed as authors will likely grow more difficult to sustain.

Furthermore, the ability to generate dynamic reimaginings of works raises questions about the ownership of the underlying data and algorithms used to produce the work, as well as the ownership of the resulting work itself.[65] These questions have important implications for the distribution and use of scholarly works, and will need to be addressed in order to ensure that the new form of scholarship is widely accessible and impactful.

With respect to ownership, a possible approach could be to recognize the underlying data and algorithms used to produce the work as separate from the resulting work itself. This distinction would allow for the distribution and use of the work to be governed by separate sets of rules and regulations, ensuring that the work remains widely accessible and impactful.

---

[63] *See Tools Such as ChatGPT Threaten Transparent Science; Here Are Our Ground Rules for Their Use*, Nature (Jan. 24, 2023),  https://www.nature.com/articles/d41586-023-00191-1.

[64] *See Thaler v. Vidal*, 43 F.4th 1207 (Fed. Cir. 2022).  *See also Thaler (Appellant) v Comptroller-General of Patents, Designs and Trademarks (Respondent)*, Case ID: 2021/0201 (United Kingdom Supreme Court, March 2, 2023).

[65] *See* Andres Guadamuz. "Artificial intelligence and copyright." *WIPO*, https://www.wipo.int/wipo_magazine/en/2017/05/article_0003.html. Accessed 3 March 2023.

In addition, the new form of scholarship raises questions about the concept of transformative work, as works are generated dynamically and in real-time, and may be significantly different from the original work in terms of content and form.[66] This raises questions about the extent to which these new works can be considered derivative works, and whether each successive derivative work - potentially *ad infinitum* - would be protected by copyright.[67]

With respect to transformative work, a possible approach could be to recognize that works generated by AI systems are transformative by nature, and to provide independent protection for each transformative work under copyright law.[68] This protection would allow for the creation and distribution of works that are significantly different from the original work in terms of content and form, while still respecting the original work and its creators.[69]  Under this practice, independent copyright protection would become as much a function of time as of content, with each successive transformative work qualifying for copyright protection independent of its ancestral works.

Finally, the new form of scholarship raises questions about compulsory licensing, as works are generated dynamically and in real-time, and may be significantly different from the original work in terms of content and form.[70] This issue raises questions about whether compulsory licensing should apply to these new works, and if so, how it should be implemented. A possible approach could be to recognize that works generated by AI systems are unique and distinct, and to provide specific exemptions from compulsory licensing for these works under copyright law.[71] Such exemptions would allow for the creation and distribution of works that are tailored to specific audiences, without being subject to the same licensing requirements as traditional works.[72] This approach would balance the need to protect the rights of authors and creators with the desire to promote innovation and the creation of new works in this exciting new field of scholarship. On the other hand, the dynamic and continuous generation of new versions of works might be easier to administer, and allow easier access to the public, under a carefully designed compulsory licensing scheme.

---

[66] *See* James Vincent, *The Scary Truth About AI Copyright is Nobody Knows What Will Happen Next*, The Verge (Nov. 15, 2022, 9:00 AM) https://www.theverge.com/23444685/generative-ai-copyright-infringement-legal-fair-use-training-data

[67] *See id.*

[68] *See id.*

[69] *See id.*

In conclusion, the proposed new form of scholarship presents substantial challenges for copyright law, raising questions concerning authorship, ownership, transformative work, and compulsory licensing. These challenges will need to be addressed in order to ensure that the new form of scholarship is widely accessible and impactful, and that it remains a powerful tool for the creation and dissemination of knowledge.

---

[70] *See* Legal Information Institute, Compulsory license, Cornell Law School, https://www.law.cornell.edu/wex/compulsory_license, (last visited Mar. 2, 2023).

[71] *See id.*

[72] *See id.*

# Building a Career in the New Form of Scholarship and Tracking Impact

The new form of scholarship proposed in this article offers a new way of creating and publishing works, one that leverages the computational power of AI systems to create more comprehensive, adaptable, and impactful works. However, it also raises questions about how academic researchers will build their careers in this new publishing genre and track their impact.

With respect to tracking impact, there will likely be new metrics and tools that emerge to help scholars track their impact across dynamic works, perhaps similar to the novel metrics provided by Altmetrics.[73] For example, metrics that measure the number of times a work has been rendered by an AI system, the number of times it has been referenced by other works, and the number of times it has been cited by other scholars could be used to track the impact of a work. Additionally, metrics that measure the engagement of users with a work, such as the amount of time spent reading a work, the number of questions asked about a work, and the number of comments or annotations or modified renderings  made about a work could also be used to track impact.

This approach to tracking impact has some similarities to the way the open source community recognizes top contributors to various projects.[74] In the open source community, contributors are recognized for their contributions based on metrics such as the number of commits they have made to various repositories, the number of bugs

---

they have fixed, and the number of new features they have added.[75] Similarly, in the new form of scholarship, scholars could be recognized for their contributions based on metrics such as the number of times their works have been rendered by an AI system, the number of times they have been referenced by other works, and the number of times they have been cited by other scholars.

These metrics could provide useful feedback for a scholar's career.  For example, if most of the citations render a work in the context of another particular work, it may be that the original authors' research may benefit from further engagement with that work. This type of information could provide valuable insights into the impact and reception of a scholar's work, helping them to refine and improve their work over time.

In conclusion, the new form of scholarship offers a new way of creating and publishing works, and raises questions about how academic researchers will build their careers in this new publishing genre and track their impact. By focusing on creating high-quality, well-researched works, and by leveraging new metrics and tools to track impact, scholars can build successful careers in this new form of scholarship.

# Potential Shortcomings of the New Form of Scholarship

While the new form of scholarship proposed in this article offers many opportunities for advancing and democratizing knowledge, it is not without its limitations and potential drawbacks.

---

[75] *See id.*

While late-binding scholarship offers many benefits, it may not be suitable for all forms of written scholarship. For example, some works may require a high degree of specificity and precision, with the specific wording of arguments being carefully crafted and polished over time. In such cases, a late-binding approach may require so much more work for the level of refinement necessary to produce a polished final product that it is necessary to sacrifice the benefits of a late-binding model to have a high quality early-bound product. Similarly, works that require extensive quantitative analysis or complex data visualization may not lend themselves well to a late-binding approach, as the dynamic nature of the format may make it difficult to present and interpret data in a clear and coherent way. In these cases, more traditional early-binding forms of scholarship may be more appropriate. Nonetheless, the potential benefits of late-binding scholarship should not be ignored, and further research and experimentation in this area may help to identify new ways in which this innovative approach to scholarship can be applied.

One concern is the issue of information overload.[76] The increased accessibility and dynamic nature of knowledge could exacerbate a current issue in academic scholarship--that of information overload. Already, there is rapid proliferation of scholarly works.[77]  The potential for large numbers of rendered editions of each scholarly work (for example, if each reading produces a separate rendering, based on ideas the reader brought to bear at the moment of engagement) could create a situation in which it becomes even more difficult for scholars and readers to sort through and make sense of the vast amount of information that is available. This could lead to a situation in which important information is overlooked, or in which scholars and readers become overwhelmed by the sheer volume of information that is available to them. However, AI support tools could also be brought to bear here, to provide interpretations and trends in how readers engaged with the work and sifting through them for the most relevant content.

---

[76] *See* David Bawden and Lyn Robinson. "Information overload: An overview." (2020).

[77] *See* Lutz Bornmann and Ruediger Mutz. "[1402.4578] Growth rates of modern science: A bibliometric analysis based on the number of publications and cited references." *arXiv*, https://arxiv.org/abs/1402.4578. Accessed 3 March 2023.

Another concern is the potential displacement of some knowledge workers, such as researchers and writers, who may no longer be needed to produce knowledge in the traditional sense.[78] Already, there are aspects of the genre of work typically completed by research assistants that are being rendered obsolete by AI writing support.[79] With the use of AI systems to create and render scholarly works, some knowledge workers may find that their skills and expertise are no longer in demand, leading to job displacement and economic insecurity.[80] Nevertheless, such displacements are common across the history of technological change, and new opportunities for employment and engagement will arise that we cannot yet envision.[81]

While the use of AI systems is likely to enhance the quality and comprehensiveness of scholarly works, there is also a concern that the quality of these works may suffer if the AI systems are not designed and used in an appropriate way. Without appropriate quality control mechanisms in place, the accuracy and reliability of scholarly works produced using AI systems may be called into question.

---

[78] *See* John Howard. Artificial intelligence: Implications for the future of work. *Am J Ind Med*. 2019; 62: 917– 926.

[79] *See* "AI research assistants and tools to optimize your workflows." *Typeset*, 29 December 2022, https://typeset.io/resources/ai-research-assistants-tools/. Accessed 3 March 2023.

[80] *See* John Howard. Artificial intelligence: Implications for the future of work. *Am J Ind Med*. 2019; 62: 917– 926.

[81] *See* Daniel Aaronson and Kenneth Housinger. "The impact of technology on displacement and reemployment." Economic Perspectives-Federal Reserve Bank of Chicago 23 (1999): 14-30.

Finally, there may be resistance to this new form of scholarship among some scholars and academic institutions, who may be hesitant to embrace a new approach that challenges traditional ways of producing and disseminating knowledge.[82] This resistance could arise from a lack of familiarity with the new form of scholarship, concerns about the potential loss of traditional knowledge work roles, identity threats, or a general skepticism about the potential of AI systems to create high-quality scholarly works.[83]

It is our hope that these challenges can be addressed through appropriate design, development, and use of AI systems, and via gradual social acceptance of AI in such roles.

# An Iterative Approach to Scholarship: Balancing Tradition and Innovation

A gradual, iterative approach may help balance tradition and innovation in order to ensure the viability of this new form of scholarship. For example, the new form of scholarship could be gradually integrated into the existing scholarly process, and in which the traditional early-binding form of scholarship is still valued and preserved for historical, archival, and attribution purposes.

---

[82] *See* Ajay Kumar Singh and Rajender Kumar. "Correlates of professional obsolescence among researchers." *Defence Science Journal* 69.6 (2019): 557-563.

[83] *See* Ekaterina Jussupow, Kai Spohrer, and Armin Heinzl. "Identity threats as a reason for resistance to artificial intelligence: survey study with medical students and professionals." *JMIR Formative Research* 6.3 (2022): e28750.

In this approach, scholars would have the opportunity to experiment with the new form of scholarship, testing and refining their ideas in real-time, and leveraging the computational power of AI systems to create more comprehensive, sophisticated, and impactful works. At the same time, they would still have the option to produce works in the traditional early-binding form, providing a stable and enduring record of their ideas and findings.

In addition, the new form of scholarship would be subject to ongoing evaluation and refinement, as scholars and copyright experts continue to explore and address the challenges raised by the new form, and as the field of AI and its applications continue to evolve.

By gradually integrating the new form of scholarship into the existing scholarly process, and by preserving and valuing the traditional form, this approach offers a path forward that is both exciting and sustainable.

# Conclusions: The Path Forward for Effective, Timely, Accessible, Democratized, and Evergreen Scholarship

The proposed new form of scholarship, with its ability to generate dynamic reimaginings of works, its potential to leverage the computational power of AI systems, and its potential to change the nature of both the scholarly process and the artifacts it produces, offers the possibility to explore powerful new directions in the creation and dissemination of knowledge. In order to realize this potential, a balanced approach that values both tradition and innovation may be helpful, as will a careful consideration and thoughtful revision of current copyright law.

The new form of scholarship offers the possibility of effective, timely, accessible, democratized, and evergreen scholarship, enabling scholars to test and refine their ideas in real-time, and to create more comprehensive, sophisticated, and impactful works. It also offers the possibility of a more dynamic and fluid process of inquiry, and the ability to generate works that are precisely tailored to the needs and interests of specific audiences, making the knowledge they contain more widely accessible and impactful. There are substantial legal questions relating to late-binding scholarship, pertaining to copyright and doctrines concerning authorship, ownership, transformative work, and compulsory licensing.  In this article, we have sought to lay initial groundwork for considering such questions and paving the way for this novel and potentially powerful new form of scholarship.

## Acknowledgments

We thank Ross Eubank for his assistance on this article. This material is based upon work supported by the National Science Foundation under Grant No. DUE-2121572.